\documentclass[aps,preprint,superscriptaddress,showpacs]{revtex4-1}

\usepackage{times}
\usepackage{amsmath,bm,amsfonts}
\usepackage{dcolumn}% Align table columns on decimal point
\usepackage{graphicx}
\usepackage{latexsym}

\usepackage{color}

\begin{document}
%\preprint

\title{Dirac cone engineering in Bi$_2$Se$_3$ thin films}

\author{Hosub \surname{Jin}}

\author{Jung-Hwan \surname{Song}}
\email[Corresponding author. Electronic address:]{jhsong@pluto.phys.northwestern.edu}

\author{Arthur J. \surname{Freeman}}

\affiliation{Department of Physics and Astronomy, Northwestern University,
Evanston, Illinois 60208, USA}

%\date{\today }

\begin{abstract}
Topological insulators are distinguished from normal insulators by their bulk insulating gap
and odd number of surface states connecting the inverted conduction and valence bands and showing
Dirac cones at the time-reversal invariant points in the Brillouin zone. Bi-based three-dimensional
strong topological insulator materials, Bi$_2$Se$_3$ and Bi$_2$Te$_2$, are known as high temperature
topological insulators for their relatively large bulk gap and have one simple Dirac cone at the $\Gamma$
point. In spite of their clear surface state Dirac cone features, the Dirac point known as a Kramers
point and the topological transport regime is located below the bulk valence band maximum. As a result of
a non-isolated Dirac point, the topological transport regime can not be acquired and there possibly exist
scattering channels between surface and bulk states as well. In this article we show that an ideal and
isolated Dirac cone is realized in a slab geometry made of Bi$_2$Se$_3$ with appropriate substitutions
of surface Se atoms. In addition to Dirac cone engineering by surface atom substitutions, we also
investigate Bi$_2$Se$_3$ thin films in terms of thickness and magnetic substitutions, which can be
linked to applications of spintronics devices.
\end{abstract}

%\pacs{Check the pacs number.}

\maketitle

%\emph{\textbf{Introduction}}
Topologically protected surface states, free of back-scattering, and spin-momentum locked massless Dirac cones
are key features of three-dimensional topological insulators~\cite{PhysRevLett.98.106803}, from which
it is possible to realize a number of new applications and new fundamental physical phenomena such as
fault-tolerant quantum computing and anomalous quantization of magnetic-electric coupling~\cite{PhysRevLett.100.096407,Xiao-LiangQi02272009}.
To utilize their distinctive properties and applications, topological insulators should be tuned in such a way
that the ideal and isolated Dirac cones, i.e., the topological transport regime, should be easily accessible
without any scattering channels. The topological transport regime can perhaps be best demonstrated by
bismuth-based compounds; a new class of topological insulators such as Bi$_2$Se$_3$ and Bi$_2$Te$_3$,
discovered by very recent theoretical and experimental studies~\cite{Xia2009,Zhang2009}, has one simple
surface state Dirac cone at the $\Gamma$ point unlike Bi$_{1-x}$Se$_x$ alloys~\cite{Hsieh2008}. However,
according to previous first-principles calculations~\cite{Xia2009,PhysRevLett.103.146401,zhang-2010},
clean surfaces of these materials have the topological transport regime located below or very close to
the bulk valence band maximum, and thus vanishing Dirac fermion density at the Dirac point cannot be
readily achieved, which is critical for manifesting novel phenomena such as an anomalous half-integer
quantization of Hall conductance. This situation also possibly introduces multiple scattering sources.
The surfaces states within the valence bands have scattering channels with bulk states~\cite{Roushan2009,gomes-2009}.
The surface states of the Dirac cone far from the Dirac point show hexagram iso-energy contours
rather than circular ones in the two-dimensional Brillouin zone, which is also a scattering channel
between the surface states~\cite{Y.L.Chen07102009,PhysRevLett.103.266801}. Apart from the scattering
sources in which only a single surface is involved, one more important scattering source, but for which
only a few studies exist~\cite{PhysRevLett.101.246807,zhang-2009}, is the inter-surface interaction
from two opposing surfaces in a very thin film geometry. The inter-surface interaction in this case
develops a small gap at the Dirac point.

Here we use a first-principles all-electron approach to investigate the effect of substitutions at the surface layer,
the inter-surface interactions, and finally magnetic dopants. We propose substitutions at the surface layer
to manipulate the level of the Dirac point relative to the valence band maximum, which may result in an isolated topological
transport regime. Tuning the Dirac cone in Bi$_2$Se$_3$ was first demonstrated experimentally by Hsieh, D. \emph{et al.}~\cite{Hsieh2009},
and our theoretical work is consistent in suggesting another way for tuning and explaining its mechanisms.
With this Dirac cone engineering and exploiting the inter-surface interaction, we design
asymmetric ultra thin films to obtain a double Rashba Hamiltonian which is an extension of the normal Rashba model,
but containing both electron and hole excitations, and thus can show the opposite sign of spin Hall conductivity.
We also demonstrate that magnetic ion substitution opens a gap in the Dirac cone of a single surface,
which gives rise to the absence of the inter-surface interaction and finally recovers a single Dirac cone
at the opposite surface in the ultra thin films. Manipulation of the Dirac cone and the inter-surface interactions will,
depending on the treatment of the surface layers proposed here, yield ideal coherent spin transports utilizing
a single surface of ultra thin films or p- and n-type spin currents in a single thin film from intrinsic
double Rashba effects. This opens new technological opportunities such as spin-polarized p-n junction devices
including electrical control of spin Hall conductivities. Recent experiments on topological insulator thin films
using molecular beam epitaxy~\cite{zhang-2009} might be useful to realize the suggested configurations here.

By appropriate substitutions at the surface, the changes in the surface potential affect
the level of surface state and finally we can acquire the ideal Dirac cone which is placed inside the bulk band
gap of Bi$_2$Se$_3$. When we substitute S and O for Se or sandwich a mono-quintuple
layer of Sb$_2$Se$_3$ at both ends of the Bi$_2$Se$_3$ film, the topological surface state
of the Dirac cone shape connecting inverted bulk conduction and valence bands moves upward.
Despite the robustness of the surface state, the position of the Dirac point and quasi-particle dispersions
depend on surface treatment sensitively. Especially for the case of O substitutions,
an ideal and isolated Dirac cone with point-like Fermi surface at $\Gamma$ is realized (Fig.1(d)-(f));
this plays a central role for investigating the intriguing properties of (2+1)
dimensional helical Dirac Fermions as well as spintronics applications.

In spite of a Dirac cone dispersion of the pristine Bi$_2$Se$_3$ thin film shown in Fig.1(a),
its Dirac point is quite close to the bulk states and both Dirac point and bulk states
appear at the same energy level, which can be a source of scattering and weaken the most
important property of the protected surface state without back-scattering. Elastic impurity scattering
between embedded surface states and bulk states was detected in recent STM and STS experiments~\cite{gomes-2009}.
Meanwhile, the linear part of the surface state inside the bulk band gap region shows a hexagram shape
of iso-energy contours near the $\Gamma$ point in the hexagonal surface Brillouin
zone, where other finite $\vec{\mathbf{q}}$ elastic scattering channels are possible.
To be more useful for device applications and fundamental research, an isolated and
circular Dirac cone which is near the Dirac point free from bulk states is required
for minimizing possible scattering channels. As shown in Fig.1(e) and 1(f), there is a circular
cross-section to the iso-energy contour in the energy range of $\sim\pm 0.1$eV above
and below the Fermi level which is free from scattering originating from the hexagram cross-section.
At the same time, electron-hole symmetry of the Dirac cone is definitely shown in this energy range.
In the finite slab geometry result, there is a small gap due to the finite size effect.
We expect that the semi-infinite Bi$_2$Se$_3$ case with O substitution at the surface can show
a single isolated Dirac cone in the energy window inside the bulk gap in experiments.
Such configurations make it easy to manipulate the systems and to see pure Dirac cone
driven effects. Due to the lack of bulk states near the Dirac point, the Fermi level pinned
at the Dirac point in the neutral system can be easily manipulated by chemical doping or gate
voltage. STS measurements on our O-substituted TI surface should not show any spatial
oscillations at certain scattering wave-vectors in the energy range inside the Bi$_2$Se$_3$
band gap. The ideal Dirac cone of the surface state is a good starting point to realize intriguing quantum
phenomena suggested by the model Hamiltonian, which emphasize the importance of the
topological transport regime. Especially, it provides an optimal condition to verify
the existence of the topological exciton condensation under the gate voltage~\cite{PhysRevLett.103.066402}.

From the charge density plots of the surface states at $\Gamma$ shown in Fig.2,
a large amount of surface state in the O substituted case resides in two Se atoms facing
at the van der Waals layer (Fig.2(d)), whereas in the other three cases the surface state
spreads through the outermost quintuple layers (Fig.2(a)-(c)). When O replaces Se,
it strongly binds with Bi, making a band insulator region within the quintuple layer,
and the surface state is pushed down to the van der Waals layers. Due to the confinement
of the surface state within the van der Waals layer, the Dirac cone moves upward and
the Dirac point appears inside the bulk band gap region.

For applications such as 2D devices, the investigation of the physical properties of thin films
of TI where two surfaces are separated by several nm are needed.
Considering a free standing slab of Bi$_2$Se$_3$ with $n$ quintuple layers where
$n$ is in the range from 1 to 6, there is a gap at the $\Gamma$ point,
and no surface state crossing of the Fermi level to connect the conduction and valence bands.
As $n$ increases, the gap decreases down to sub-meV at $n=6$, in agreement
with previous work~\cite{liu-2009}. The origin of these small gaps is the interaction between
surface states from opposite surfaces of the slab. For semi-infinite Bi$_2$Se$_3$, the lack
of back-scattering in the surface state is guaranteed by time-reversal symmetry, which means
that there is no scattering channel between $\vec{\mathbf{k}}$ and $-\vec{\mathbf{k}}$ surface states with
opposite spin components. For the finite TI slab with an even number of surface states from two different
surfaces, however, scattering between the two surface states from the opposing surfaces with
the opposite momentum but the same spin is allowed which creates a small gap
at the crossing point even though  the scattering between the Kramers pair at the
same surface is still forbidden~\cite{PhysRevLett.95.226801}.

Based on the fact that substitutions of the surface anion change the energy level
of the surface state Dirac cone, we designed an asymmetric ultra-thin TI slab by replacing atoms
at the end of one surface to generate an asymmetric potential and to break the
degeneracy between surface states from each side. By substitution of S, and O for Se in 3 quintuple
Bi$_2$Se$_3$ layers, we acquired two different surface states whose Dirac points
at $\Gamma$ do not coincide. There still exist inter-surface interactions
between channels with the same spin from opposing sides, but the gap opens slightly
off $\Gamma$ because the crossing points between the scattering
channels are shifted due to the asymmetric environment. As a result of the asymmetric potential
and inter-surface interaction, a gap is opened at those crossing points, but two gapless
Dirac points from asymmetric surfaces are placed at different energy levels at $\Gamma$.

The asymmetric thin film has the advantage of carrying finite spin Hall currents whose
magnitude and direction are determined by the position of the chemical potential. For
symmetric geometries, even though each topologically protected surface state carries spin Hall
currents, they are exactly canceled out by those from the opposing sides of the film.
However, in the case of an asymmetric slab, we can expect a finite spin Hall conductivity
due to imperfect cancellation. Starting from two surface states described by a helical Dirac
Hamiltonian and asymmetric potential $V_a$, the model Hamiltonian $H$ can be built
to consider the inter-surface interactions $V_I$.
\begin{equation}
H=\left[\frac{V_a}{2}\sigma_0 + v_F \vec{\sigma}\cdot(\hat{z}\times\vec{p}) \right]
\tau_3 + V_I\sigma_0 \tau_1,
\end{equation}
where $\sigma_0$ is the $2\times2$ unit matrix, and $\sigma_i$ and $\tau_i$ are Pauli
matrices indicating spin and surfaces respectively. By using an unitary transformation~\cite{winkler03}
to fold back the off-diagonal component $V_I$ and expanding up to $(V_I/V_a)^2$, $H$
can be written as
\begin{equation}
\widetilde{H}=\textrm{e}^{-S}H\textrm{e}^S=\left[\left(\frac{\widetilde{V}_a}{2}+
\frac{p^2}{2\widetilde{m}} \right) \sigma_0 + \widetilde{v}_F \vec{\sigma}\cdot
(\hat{z}\times\vec{p}) \right]\tau_3,
\end{equation}
which is exactly mapped into the Rashba Hamiltonian.

Considering the band dispersions of the surface state induced by inter-surface interactions,
and helical spin textures in momentum space (Fig.3(d)), the band dispersions are
adiabatically connected to the Rashba Hamiltonian, so that they give a non-zero
spin Hall conductance~\cite{PhysRevLett.92.126603}. More importantly, $\widetilde{H}$ is
composed of a double Rashba model where the same Hamiltonian is applied both for electron and
hole excitations. Due to the characteristics of electron and hole excitations for surface state bands
above and below the gap, the sign of the spin Hall conductivity is changed when the Fermi
level is shifted across the gap. Therefore, the direction of spin Hall currents can be
easily reversed by adjusting the chemical potential.

A way to avoid inter-surface interaction and to recover a gapless surface state in finite TI
thin films is to destroy one surface state by exerting a local magnetic field only limited
to one surface and breaking local time-reversal symmetry which makes a gap at the
Dirac point of the surface state. Due to gapping one surface state, the other surface state is free from scattering
at the Dirac point and has gapless dispersions even in thinner Bi$_2$Se$_3$ films
where there are inter-surface gaps at the Dirac point. To generate a magnetic field
localized on one surface of the Bi$_2$Se$_3$ thin film, we set up an asymmetric slab with
substitution of the end Bi-Se sub-layer by Mn and I (MnI). With MnI replacement, Mn has a 3$d^5$ configuration
where we can make use of its large magnetic moment and exchange splitting to open
a gap at the Dirac point at one surface state. To calculate this system, we adopted the LDA+$U$ scheme
containing spin-orbit interaction with $U=4.0$ eV and $J=1.0$ eV for the Mn 3$d$ orbitals.
With this stoichiometric substitution, the calculated magnetic moments are nearly 5 bohr magnetons
per Mn regardless of $U$. More importantly, MnI substitution not only generates a local magnetic field,
but also plays the role of an asymmetric potential to shift the surface state upward, similar to the case of O
substitution. Even though a single gapless Dirac cone in the Bi$_2$Se$_3$ thin film under
local magnetic fields is realized, its Dirac point is not isolated from other states.
To obtain an isolated single Dirac cone through the whole Brillouin zone in a certain
energy window inside not only the bulk gap but also the surface state gap generated by a local field,
we set up a 3 quintuple layer Bi$_2$Se$_3$ slab with MnI and O substitution for each surface.
By O substitution, the energy level of a gapless surface state moves upward and is just located inside
the gap of the other surface state opened by the local magnetic field. As shown in Fig.4(c), we can identify
two different surface states, where one is gapped by 69 meV at $\Gamma$ and the other is gapless.
The energy window of an isolated Dirac cone is limited by the gap of the MnI substituted surface state.
Making thicker magnetic layers by (MnI)$_2$ substitution, the surface state gap induced by the magnetic field grows
(Fig.4(d)). Since it is preferable to have a large energy window for practical use, the possibilities of
enlarging the gap by utilizing different configurations with different ferromagnetic materials
should be investigated in the future.

One important implication of destroying one surface state and avoiding the inter-surface interaction is
that we can make an ultra-thin film which preserves a gapless and isolated surface state.

Isolated and ideal Dirac cone surface states, protected by time-reversal symmetry,
which we have shown in this article will play an important role in charge and spin transport
experiments with their long coherent length and high mobility. Dirac cone engineering done
by the substitutions of surface atoms which alters the level of surface states without changing
the bulk states can also be applicable to another promising 3-dimensional topological
insulator, Bi$_2$Te$_3$. Asymmetric substitutions with magnetic or non-magnetic ions in the Bi$_2$Se$_3$
narrow slab showed various applicability by using the helical Dirac fermions of the surface states and
the interactions between them which are inevitable in thin film geometries of the topological insulators.

\begin{acknowledgments}
Financial support from the U.S. DOE under Grant No.DE-FG02-88ER45372 is gratefully acknowledged.
\end{acknowledgments}

%\cite{*}

\bibliography{Bi2Se3}% Produces the bibliography via BibTeX.

\begin{figure}[b]
 \centering\includegraphics[width=15cm]{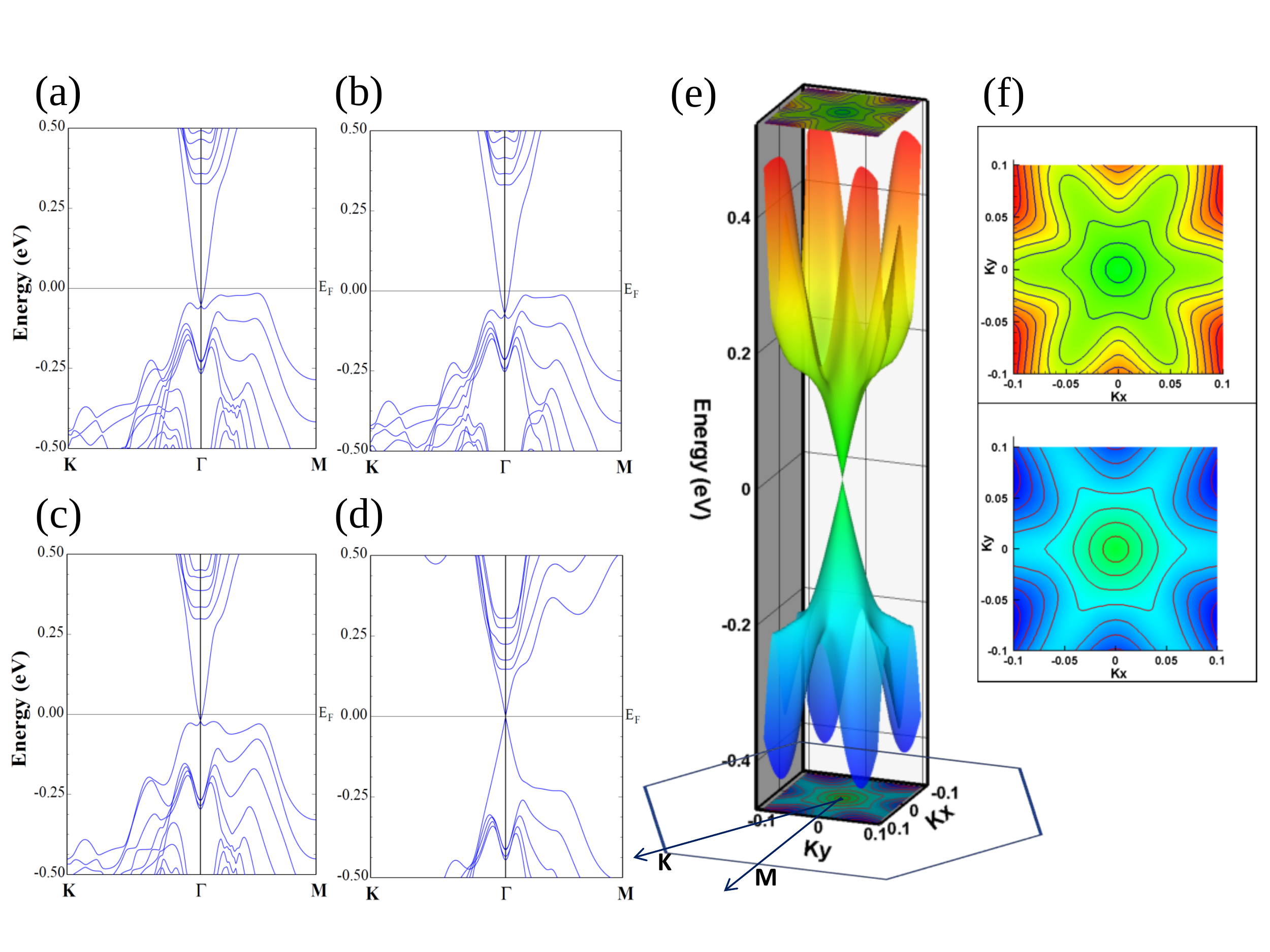}
 \caption{Quasi-particle dispersions of (a) Bi$_2$Se$_3$ pristine slab,
 (b) Bi$_2$Se$_3$ slab with S substitution of end Se atom, (c) Bi$_2$Se$_3$ slab
 sandwiched by Sb$_2$Se$_3$ mono-quintuple layer, and (d) Bi$_2$Se$_3$ slab with
 O substitution. All film geometries consist of 6 quintuple layers.
 (e) 3-dimensional plot of the surface state from O-substituted case in (d), and
 (f) its contour plot above and below the Dirac point.} \label{fig:1}
\end{figure}

\begin{figure}[t]
 \centering\includegraphics[width=15cm]{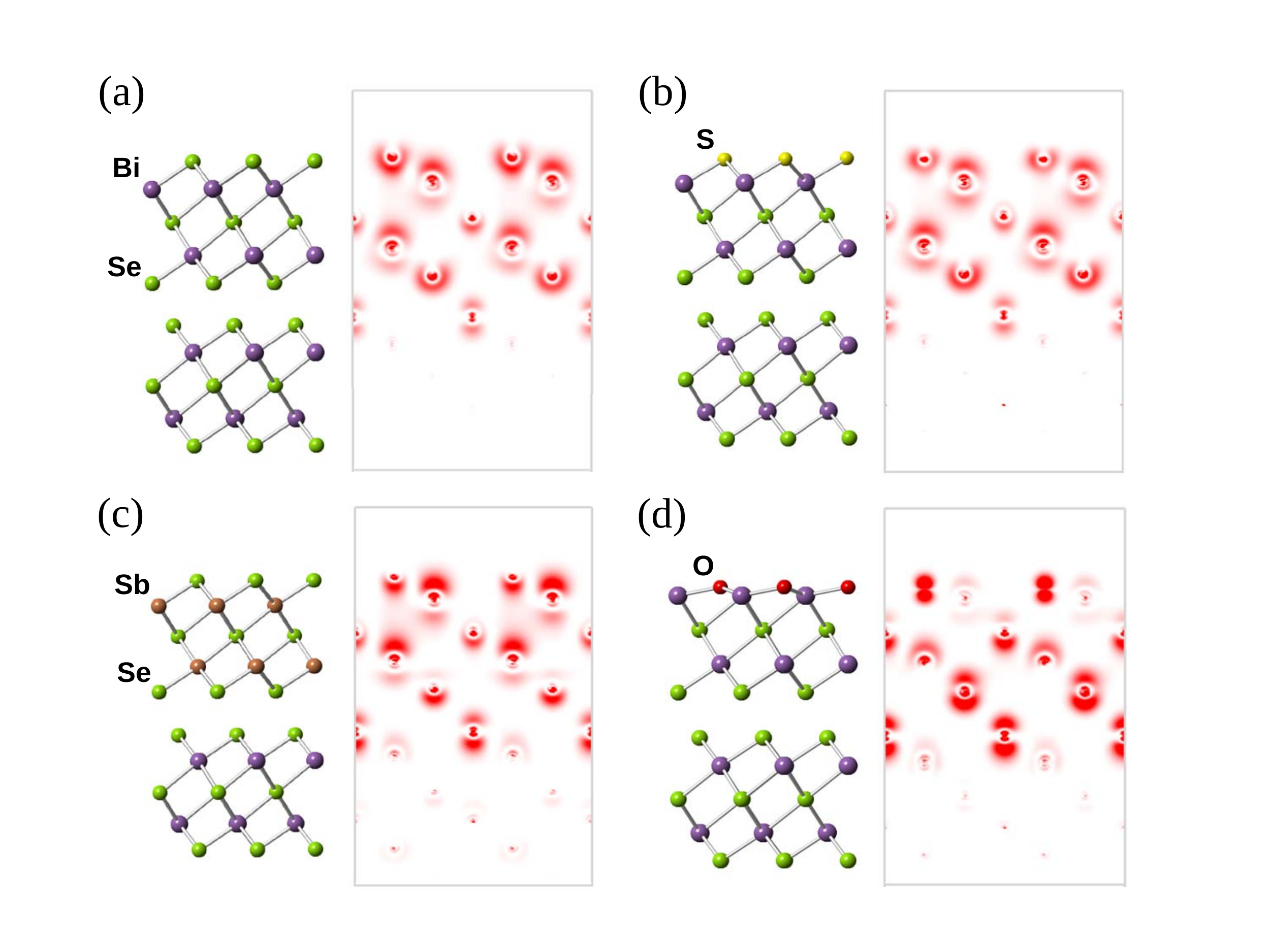}
 \caption{Charge density plots of the surface states at $\Gamma$
 from (a) Bi$_2$Se$_3$ pristine slab, (b) with S substitution, (c) with
 Sb$_2$Se$_3$ mono-quintuple layer sandwich, and (d) with O substitution.} \label{fig:2}
\end{figure}

\begin{figure}[t]
 \centering\includegraphics[width=15cm]{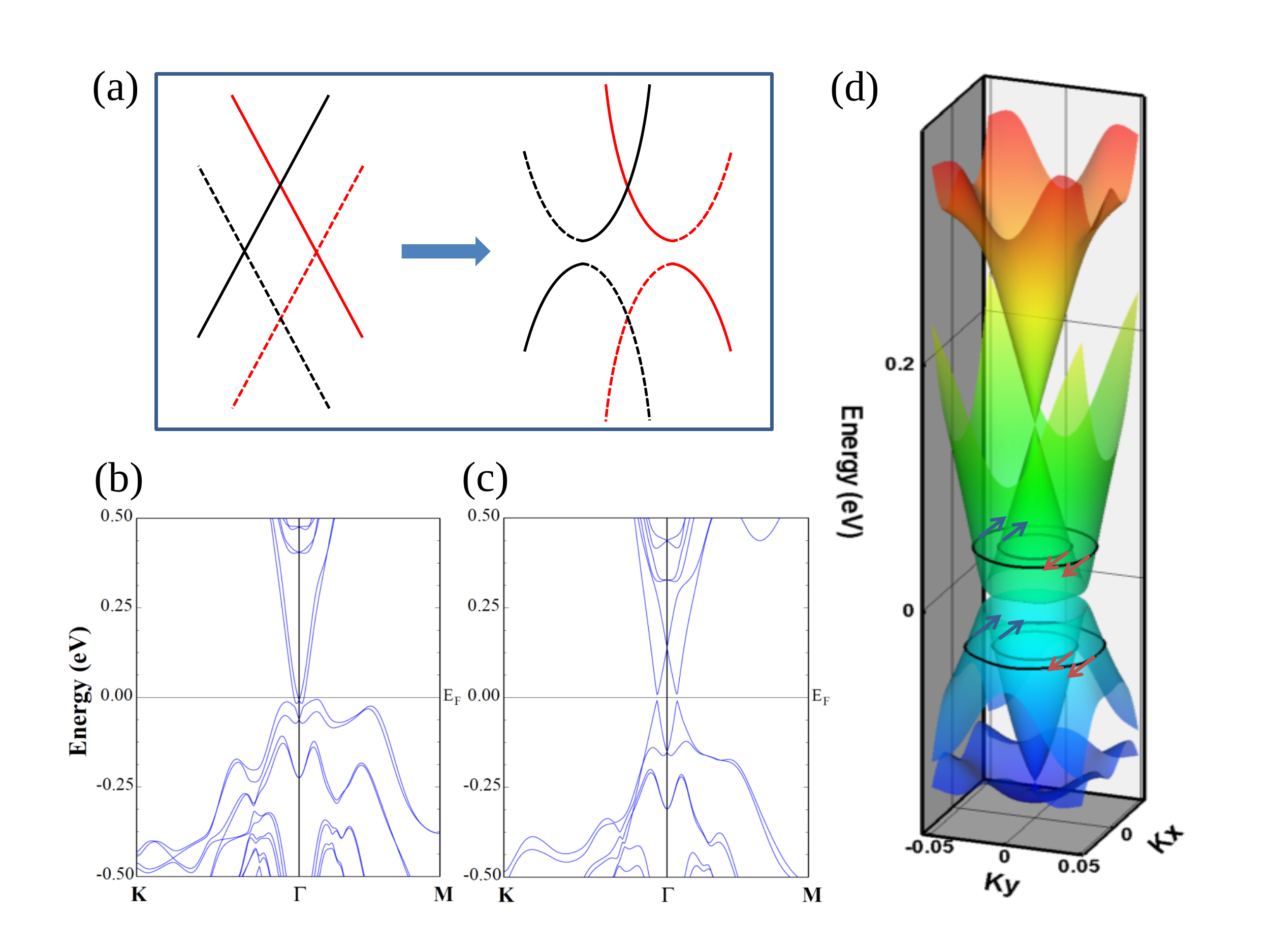}
 \caption{(a) Schematic diagram of band dispersions of surface states from two asymmetric
 surfaces, describing the changes from two independent Dirac cones to the double Rashba model
 considering inter-surface interaction. Solid and dotted lines denote two different surfaces
 and blue and red colors two spin states. Electronic band structures of three quintuple layers
 slab consisting of (b) 2 Bi$_2$Se$_3$ quintuple layers and mono-layer Sb$_2$Se$_3$, and
 (c) Bi$_2$Se$_3$ slab with end O substitution. (d) 3-dimensional plots of band structure in (c).
 Considering the spin directions in momentum space (denoted as arrows), two asymmetric surface
 states with inter-surface interaction adiabatically connect to the double Rashba model.} \label{fig:3}
\end{figure}

\begin{figure}[t]
 \centering\includegraphics[width=15cm]{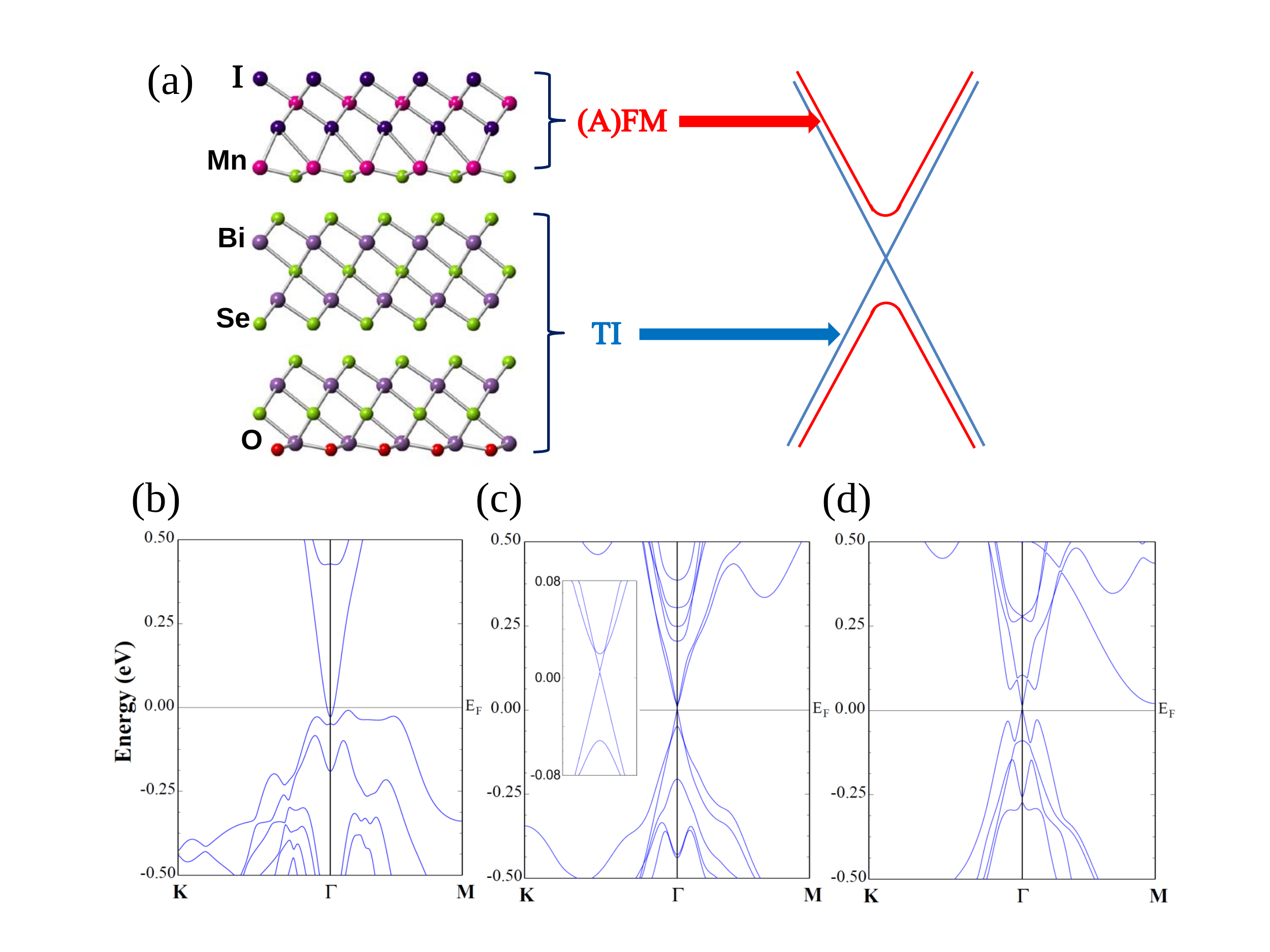}
 \caption{Crystal structure of 3 quintuple layers of Bi$_2$Se$_3$ with O substitution at one end
 and (MnI) substitution at the other end. Schematic band dispersions comparing two different
 surface states from O-substituted side (blue, gapless) and from (MnI)-substituted side (red,
 finite gap). Band structures of 3 quintuple layers of (b) pristine Bi$_2$Se$_3$ slab, (c)
 O and (MnI) substitutions at both ends, and (d) O and (MnI)$_2$ substitutions.} \label{fig:4}
\end{figure}

\end{document}